\documentclass[letterpaper,twocolumn,10pt]{article}
\usepackage{usenix2019_v3}

\usepackage{tikz}
\usepackage{amsmath}
\usepackage{amssymb}

\usepackage{filecontents}

\usepackage{subcaption}
\usepackage{algorithm}
\usepackage[noend]{algpseudocode}

\begin{document}
%-------------------------------------------------------------------------------

%don't want date printed
\date{}

\title{\Large \bf Learning to Wait:\\
  Wi-Fi Contention Control using Load-based Predictions}

%for single author (just remove % characters)
%\author{
%{\rm Your N.\ Here}\\
%Your Institution
%\and
%{\rm Second Name}\\
%Second Institution
% copy the following lines to add more authors
% \and
% {\rm Name}\\
%Name Institution
%} % end author

\author{
{\rm Thomas Sandholm}\\
CableLabs
\and
{\rm Bernardo Huberman}\\
CableLabs
\and
{\rm Belal Hamzeh}\\
CableLabs
\and
{\rm Scott Clearwater}\\
Clearway
% copy the following lines to add more authors
% \and
% {\rm Name}\\
%Name Institution
} % end author

\maketitle

\begin{abstract}
We propose and experimentally evaluate a 
novel method that dynamically changes the contention window of
access points based on system load to improve performance in a 
dense Wi-Fi deployment. 
A key feature
is that no MAC protocol changes,
nor client side modifications are 
needed to deploy the solution.
We show
that setting an optimal contention
window can lead to throughput and latency improvements
up to 155\%, and 50\%, respectively.
Furthermore, we devise an online learning method
that efficiently finds the optimal contention
window with minimal training data, and yields
an average improvement in throughput of 53-55\% during
congested periods for a real traffic-volume workload replay
in a Wi-Fi test-bed.
\end{abstract}

\section{Introduction}\label{sec:introduction}
As more and more devices come on-line and demand an increasing
amount of throughput and low latency~\cite{sandvine2016}, contention for
the available wireless spectrum has become a growing problem~\cite{ovum2019}.

Wi-Fi uses a Listen-Before-Talk (LBT) mechanism to access
unlicensed spectrum in a fair and efficient way.  The most used
algorithm in today's networks is the Binary-Exponential-Backoff (BEB)
mechanism that forces transmitters to double their wait time after
each failed transmission. Many researchers have shown in the past that
this mechanism scales poorly as the number of interfering transmitters
increases and that adjusting the contention window can improve performance~\cite{cil2012,deng2016,bianchi2000}.

The problem has been recently exacerbated by the trend in the 
latest Wi-Fi specifications~\cite{perahia2013,khorov2018} to 
bond existing 20Mhz wide channels to 40, 80, and 160Mhz bands, in order to
allow wireless end-user devices to take advantage of throughput 
increases in high-speed broadband backhauls~\cite{bennett2013}.

Setting different QoS levels on different packets helps some high
priority streams, but what if all streams are marked high priority?
In other words, what if we want to improve the Wi-Fi experience
without compromising fairness?

Many algorithms that improve BEB both in terms of throughput and
fairness have been proposed~\cite{syed2016,tian2005,xia2006,heusse2005,patras2010}, but failed to make an impact 
due to the inertia of (MAC) protocol adoption in Wi-Fi~\footnote{by design for backwards-compatibility}, in particular
for end-user devices. Furthermore, a typical dense Wi-Fi deployment exhibits 
highly complex performance dynamics that are hard to reproduce
in simulations alone. The complexity not only makes simulation results less reliable,
but may also call for more flexible, and adaptable optimization models and methods.   

In response to these challenges we have developed a novel machine-learning method
to control the Wi-Fi contention window
on access points (APs) deployed in a dense environment. Our method does not require
any changes on client devices, or to the Wi-Fi protocols, while
still being fair across both participating and exogenous devices.

We evaluate our method experimentally using a test-bed with off-the-shelf APs and
Wi-Fi stations and using a trace of traffic volumes recorded from a real
broadband deployment.

Our work extends the existing body of work
by taking both transmitters and system load into account to predict
the optimal wait times at any given time, and using a trained model capable of
capturing correlations between easily observed measurements and 
optimal contention windows.

The key contribution is threefold:
\begin{itemize}
	\item{we propose a model to predict optimal contention windows (Section~\ref{sec:model}),}
\item{we explore and quantify the opportunity
of setting an optimal contention window experimentally in a test-bed (Section~\ref{sec:contention}), and}
\item{we propose a learning algorithm to continuously train
our predictive model to adjust the contention window based on system load (Section~\ref{sec:prediction})} 
\end{itemize}

The rest of the paper is organized as follows. We introduce the fundamentals of the Wi-Fi backoff mechanism in
Section~\ref{sec:background}. Then we present our predictor models in~\ref{sec:model},
and the dataset used to replay
traffic into our test-bed (Section~\ref{sec:dataset}), which in turn is described
in Section~\ref{sec:testbed}. In Section~\ref{sec:contention} we run a series
of experiments to showcase the opportunity in improving the contention mechanism
implemented in current networks (BEB). Section~\ref{sec:prediction} then shows
the results of training and applying a predictive model for contention window control.
In Section~\ref{sec:implementation} we describe the architectural components
of our implementation.  We discuss related work in
Section~\ref{sec:relatedwork}, and finally we conclude in Section~\ref{sec:conclusion}.

\section{Background}\label{sec:background}
Shared medium access in Wi-Fi is implemented by the distributed
coordination function (DCF) in the medium access control (MAC) layer.
It is a contention-based protocol based on the more general
carrier sense multiple access with collision-avoidance (CSMA/CA)
protocol. When a frame is to be transmitted, a Listen-Before-Talk (LBT) 
mechanism is employed where the channel is first sensed and if busy 
the transmission attempt is delayed for a backoff period. The backoff
period is determined by picking a uniformly random number of wait time slots, say $r$, in the interval $[0..CW]$,
where $CW$ is the contention window. The starting point $CW$ is referred to as $CW_{min}$.
The new transmission attempt is then made after $r$ time slots.
Each slot is a standardized time interval, typically about $9\mu s$.
The most commonly used backoff algorithm, binary exponential backoff (BEB), doubles the contention
window, $CW$, for every failed attempt, up to a maximum value of $CW_{max}$. After a given
number of failed attempts the frame is dropped, and the error is propagated to higher-level protocols
or the application. In case of a successful transmission the contention window is reset to $CW_{min}$.

The enhanced distributed channel access (EDCA) amendment (802.11e)~\cite{mangold2002} introduced a QoS extension to
DCF~\footnote{it is therefore also sometimes referred to as EDCF (extended distributed coordination function)} for 
contention-based access. 
A number of parameters can be configured in different service classes, or transmission queues
to give different priorities to different flows. One set of these parameters is the $(CW_{min},CW_{max})$ tuple.
The standard specifies default values for this tuple, but access point administrators may
change these values at will. For best effort (BE) traffic the default values are $(15,63)$~\footnote{802.11e has $(15,1023)$ as default, but the hostapd implementation we use set $(15,63)$ as default, which is why we use it here. We test the wider range too in Section~\ref{sec:windowrange}}.   
Our approach is based on changing these parameters dynamically and always setting $(CW_{min},CW_{max})$
to $(CW_{opt},CW_{opt})$,
to effectively disable the exponential backoff, and leave it fully in our control how long the
transmitters have to wait on average before attempting to retransmit ($9CW_{opt}/2\mu s$~\footnote{ignoring
frame size variations, time of sensing before starting the backoff process, and time waiting for a transmission ACK,
which are all independent of the backoff algorithm and CW settings}).

Now the problem is reduced to finding and setting the optimal contention window, $CW_{opt}$ that maximizes
some QoS parameter, such as throughput or latency. 
Here, we consider only single-step prediction of $CW_{opt}$. That is, we observe some state of the system
in period $t_0$ and then set the contention window to use in period $t_1$. The motivation behind this setup
is that predictions further into the future will be less accurate. Furthermore, if there is a long gap between
observation and enforcement, and between enforcements, the true $CW_{opt}$ may change within allocation periods,
leading to suboptimal allocations.      

Note, that this rules out machine learning techniques such as reinforcement learning (RL), where 
allocation decisions in the current period is assumed to impact rewards in multiple future periods.  
In other words, the greedy optimization strategy is always optimal in our case. However,
as we will discuss later, our approach is inspired by and borrows some techniques from RL.

\section{Model}\label{sec:model}
In~\cite{syed2016} an adaptive backoff algorithm (ABA) is
derived from the probability of collisions, given
a fixed number of active transmitters, and a configured minimum CW, $CW_{MIN}$.
The optimal CW is then estimated as follows:
\begin{equation}
 CW_{opt} = \frac{CW_{MIN}}{2}\times a-1
\end{equation}
where $CW_{MIN}$ is the default EDCA minimum contention
window for the service class (15 for best effort), and $a$
the number of active transmitters~\footnote{Note this formula 
assumes that $CW_{MIN} \ge 2$ and that $a \ge 2$. If $a < 2$
the collision probability is $0$ and the $CW$ backoff process
never commences}.

We propose a generalization of the ABA model as follows:
\begin{equation}\label{eq:poly}
	log(CW_{opt}) = \theta_0 + \theta_1 a + \theta_2 {t\!p}
\end{equation}
where the $\theta$ comprise the model coefficients to be trained, $a$ the observed number of active APs, and
$t\!p$ the observed aggregate throughput from the last period.

Apart from the generalization and the additional load term, the model
allows for continuous adaptation of the coefficients to fit the observed
data. We will discuss a learning method that trains this model
online in Section~\ref{sec:prediction}.

The load term was introduced to account for environment interference
impacting the throughput, i.e. factors beyond the APs that we control.

Given the intended use for training with Machine-Learning methods, we call
this general model (Equation~\ref{eq:poly}) the Machine-Learning Backoff Algorithm ({\bf MLBA}) model.

The log transform, which implies that the predictors are multiplicative
as opposed to additive, is motivated by experiments
in Section~\ref{sec:prediction}. 
Intuitively, the 
expected throughput of a transmitter at any given time is inversely proportional to the current contention window,
and is also proportional to the product of the probabilities that other transmitters will not transmit at that time
(see Appendix~\ref{sec:heuristics}).

\subsection{Model Training}\label{sec:parameter_estimation}
\subsubsection{MLBA-LR}
The simplest way to estimate the coefficients for an observed set of $\{a,{t\!p}\}$ 
input parameters (predictors) and optimal CW (response) is by Least squares regression, e.g. ordinary least squares (OLS),
where a coefficient is estimated by the covariance of the parameter with the response variable (output).
So to estimate $\theta$ above we compute:
\begin{align*}
	\hat{\theta}_0 &= log(CW_{opt}) - \hat{\theta}_1 a - \hat{\theta}_2 {t\!p} \\  
	\hat{\theta}_1 &= \frac{Cov[a,log(CW_{opt})]}{Var[a]}  \\
	\hat{\theta}_2 &= \frac{Cov[{t\!p},log(CW_{opt})]}{Var[{t\!p}]} 
\end{align*}
We call the backoff algorithm deploying this form of parameter estimation {\bf MLBA-LR}.

\subsubsection{MLBA-NB}
Another approach is to learn model parameters through a {\it Naive-Bayes} method, where
the contention window that maximizes the conditional probability of a given state $s=\{a,{t\!p\}}$
is chosen. The probability can be computed according to Bayes theorem as:
\begin{equation}\label{eq:nb_eq}
	CW_{opt} = \operatorname*{arg\,min}_{CW} \left\{P(CW|s) \triangleq \frac{P(s|CW)\times P(CW)}{P(s)}\right\}
\end{equation}
We call the backoff algorithm deploying this form of contention window estimation {\bf MLBA-NB}.

\subsubsection{MLBA-DNN}
As a final model we propose a {\it Deep Neural Network} to estimate an optimal
contention window $CW$ given a state $s$. 
A Deep Neural Network (DNN) is defined as having multiple hidden layers.
Between every two layers is a (nonlinear) activation function that determines how to map output from one layer into
inputs of the next layer. Our MLB-LR model can hence be seen as a collapsed single-layer DNN.
Our minimal network configuration is, 
one input layer, two hidden layers,
and one output layer. The output layer renders the final prediction and is thus
often single-dimensional.
The 3-layer (2 hidden layer) DNN model can then be expressed as:
\begin{equation}\label{eq:dnn_eq}
	CW_{o\!pt}(x) = \mathbf{b}^{(3)}+ \mathbf{w}^T h(\mathbf{b}^{(2)} + \mathbf{W}^{(2)} h(\mathbf{b}^{(1)} + \mathbf{W}^{(1)}\mathbf{x}))
\end{equation}
where $h()$ is the hidden layer activation function, in this case we use rectified linear units~\cite{hahnloser2000}, $ReLu(x) = max(x,0)$, 
$\mathbf{x}$ is the input vector of $\{a,t\!p\}$ tuples, $\mathbf{W}^{(k)}$ is the matrix of weights
for hidden layer $k$, $\mathbf{b}^{(k)}$ is the bias vector for layer $k$ ($k=3$ is the output layer)
, and $\mathbf{w}$ is the vector of weights between the last hidden layer and the single cell
output layer.

To train this model (using backpropagation) we apply mean-square-error as loss function~\footnote{recommended for regression tasks}, 
and the {\it adam}~\cite{kingma2014} stochastic gradient descent algorithm. 
We have flexibility in selecting the number of nodes in the hidden layers; more nodes would mean a more accurate
fit but longer training time. It also depends on the variance of the number of transmitters and load. 
A high variance and complex interrelationships may require more nodes to model appropriately.
We call the backoff algorithm deploying this form of contention window estimation {\bf MLBA-DNN}.

Note here that a typical DNN is a supervised learning model, but as we shall see in Section~\ref{sec:prediction},
we create the training data on the fly and hence our method becomes unsupervised.

\section{Traffic Volume Data}\label{sec:dataset}
Given that the model we propose is predictive and should be capable of learning some hidden
behavior in traffic volume dynamics, we collect a real-world trace from
a residential deployment with cable modems connected to a cable headend (CMTS)~\footnote{cable modem termination system}
over a HFC~\footnote{Hybrid fiber-coaxial} network. The data comprise download volumes on a per-second basis
for each cable modem. Volumes were captured on July 1st 2017 from 8 cable modems.

Due to the fact that the rates are set dynamically in a Wi-Fi network based on
the measured signal-to-interference-plus-noise ratio (SINR), we only capture 
whether the modem is active or not (the sum of active modems is quantity $a$ in Section~\ref{sec:model}), i.e. transmitting
any data in each second of the trace, and we then let our test-bed (see Section~\ref{sec:testbed}) find the best rate.

\begin{figure*}
	\centering
	\begin{subfigure}[b]{0.49\textwidth}
		\includegraphics[width=\textwidth]{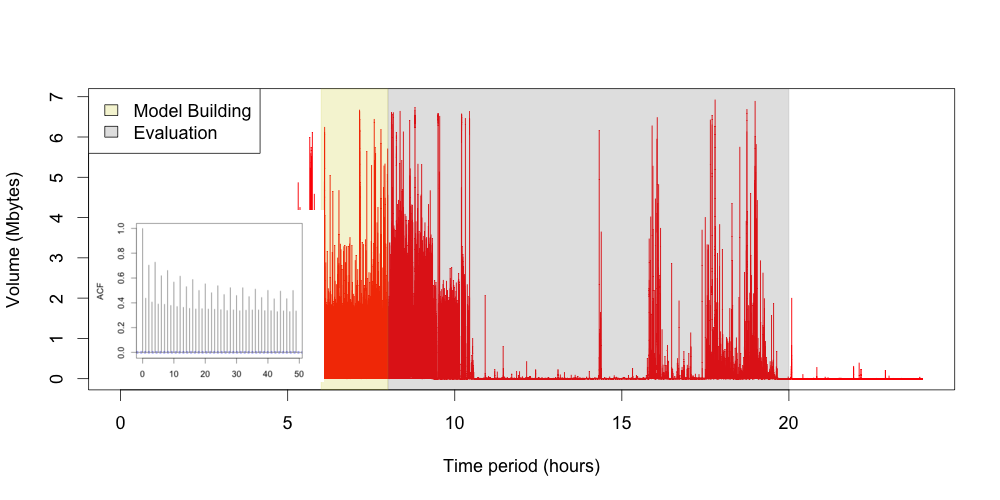}
	\end{subfigure}
	\begin{subfigure}[b]{0.49\textwidth}
		\includegraphics[width=\textwidth]{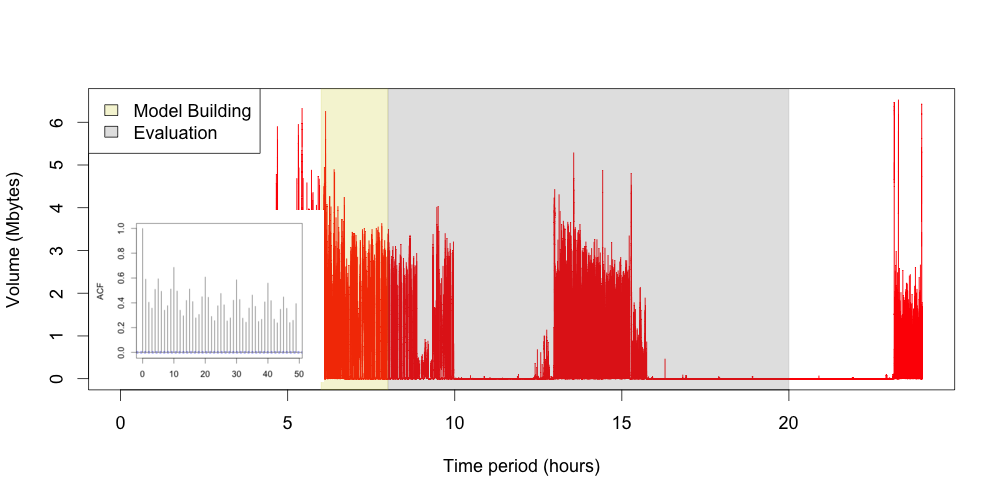}
	\end{subfigure}
	\begin{subfigure}[b]{0.49\textwidth}
		\includegraphics[width=\textwidth]{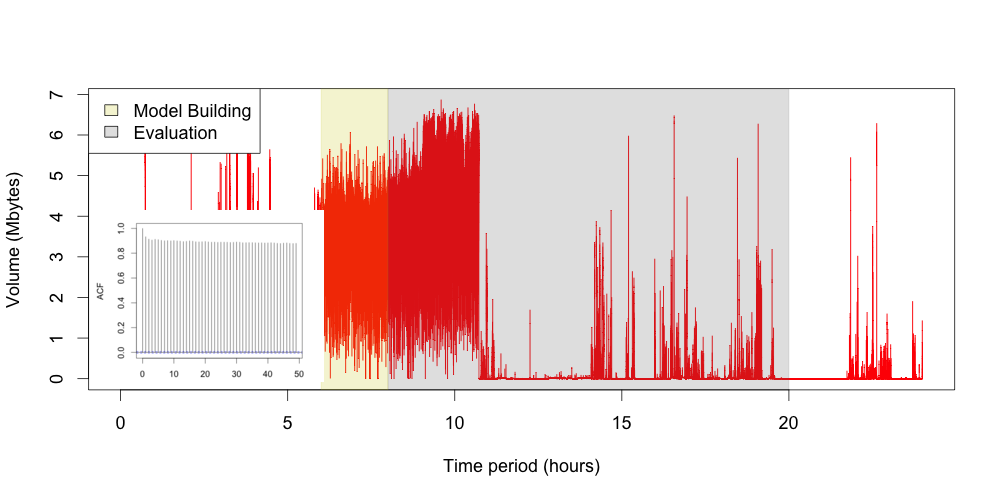}
	\end{subfigure}
	\begin{subfigure}[b]{0.49\textwidth}
		\includegraphics[width=\textwidth]{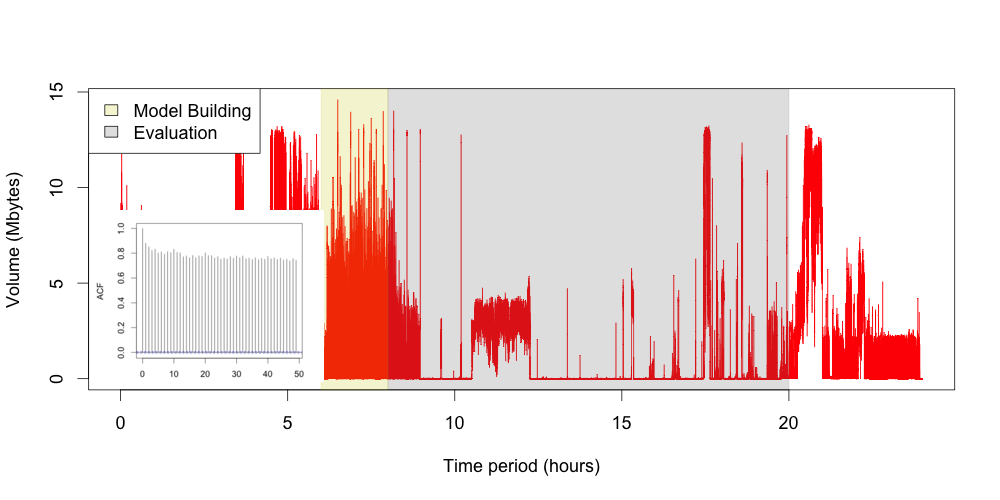}
	\end{subfigure}

	\begin{subfigure}[b]{0.49\textwidth}
		\includegraphics[width=\textwidth]{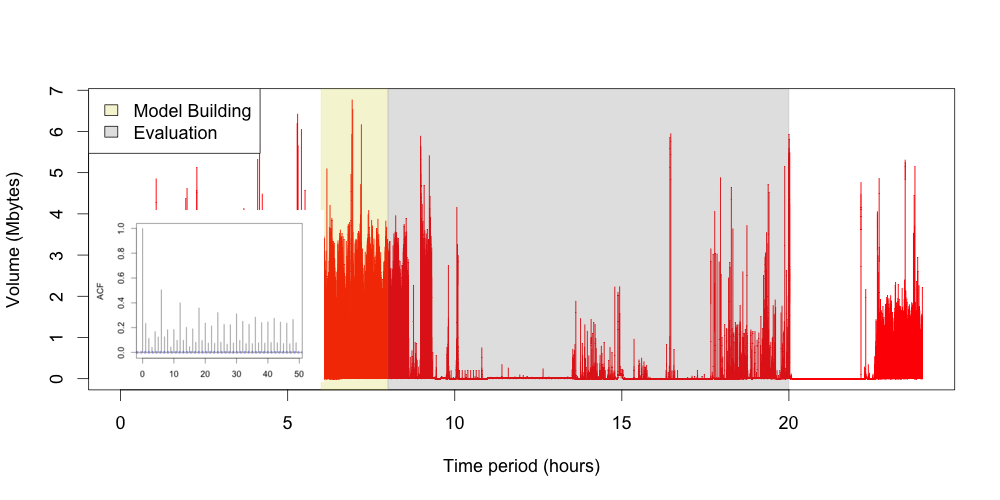}
	\end{subfigure}
	\begin{subfigure}[b]{0.49\textwidth}
		\includegraphics[width=\textwidth]{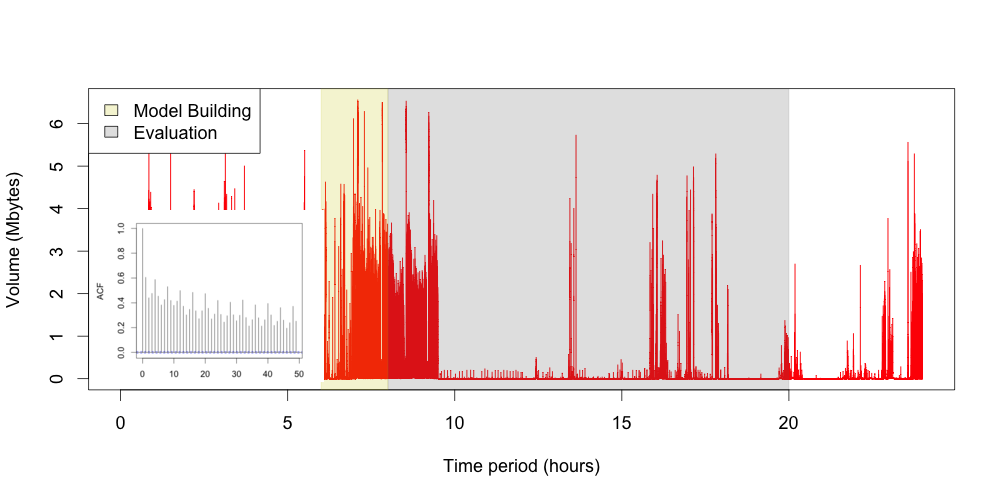}
	\end{subfigure}
	\begin{subfigure}[b]{0.49\textwidth}
		\includegraphics[width=\textwidth]{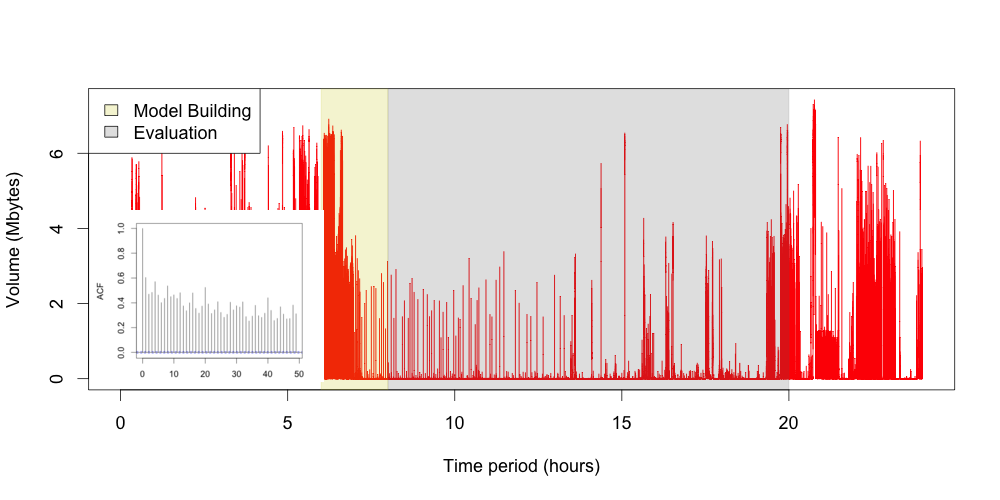}
	\end{subfigure}
	\begin{subfigure}[b]{0.49\textwidth}
		\includegraphics[width=\textwidth]{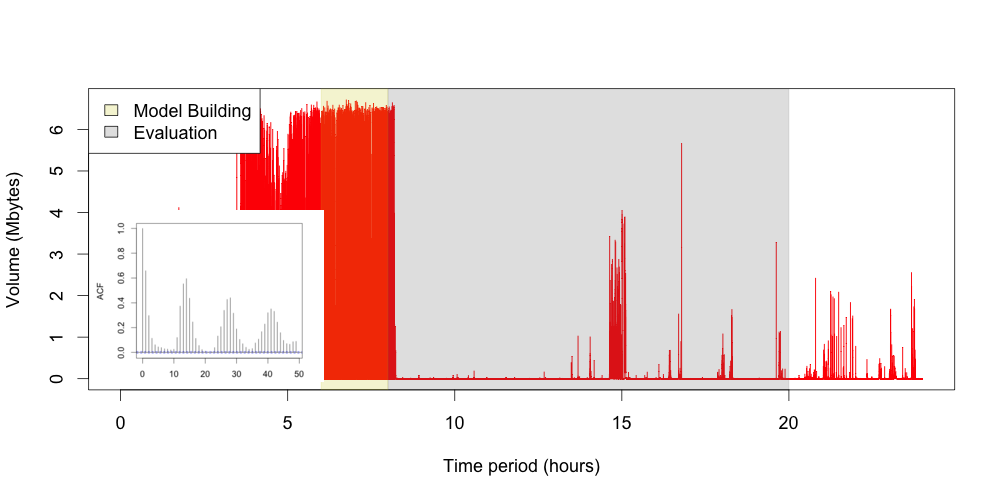}
	\end{subfigure}
	\caption{Workloads with ACF insets (seconds) from 8 cable modems}
	\label{workloadstreams}
\end{figure*}

The individual traces can be seen in Figure~\ref{workloadstreams}. We selected one hour for model building analysis, one hour for 
meta parameter evaluation and five hours for model prediction evaluation (see Section~\ref{sec:prediction}).

\section{Testbed}\label{sec:testbed}
Our test-bed comprises 8 Wi-Fi Access Points transmitting on
the U-NII-3 80Mhz band.
Two APs are TP-Link Archer AC1750 routers, and 6 are GLi GL-AR750S-Ext devices. 
All run the latest OpenWrt release with a patch we developed to control $CW_{min}$ and
$CW_{max}$ from the hostapd control interface and CLI (see more details in Section~\ref{sec:implementation}).

We use an additional 8 GLi GL-AR750S-Ext as Wi-Fi clients, and 8 Raspberry Pi Model 3B
to run iperf3 servers that the clients connect to. The iperf servers could also
run on the APs but it leads to CPU contention, limiting the throughput instead of
airtime, at higher rates. 

All transmissions are done with 802.11ac VHT and a theoretical max throughput rate
of 433Mbps. With iperf3 servers running on the APs the max throughput is around 100Mbps,
whereas separating out the servers on the Pis improved the throughput to about 300Mbps.

All traffic is going in the direction from the AP to the clients, with unrestricted~\footnote{bit rates are determined by the signal quality} TCP flows.

The distance between the APs and the clients vary between 5 and 10 feet, and the distance
between the APs vary between 2 and 6 feet. Finally, the distance between the clients
vary between 2 to 5 feet. Both clients and APs are stationary throughout all experiments.

\begin{figure}[htbp]
        \centerline{\includegraphics[scale=0.25]{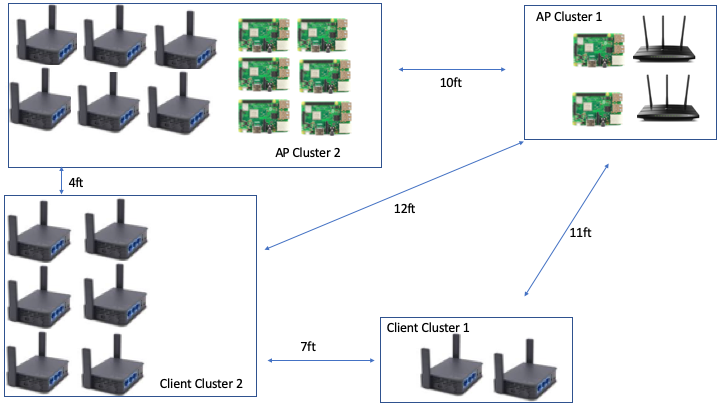}}
        \caption{Test-bed layout sketch (not to scale). Note icon pictures show devices used, and
	AP cluster 2 uses the same device as the clients (GLi).}
\label{testbed}
\end{figure}

A room layout sketch of the physical setup is shown in Figure~\ref{testbed}.

All APs and clients are connected to an Ethernet switch to avoid control and measurement traffic
interfering with the Wi-Fi link. The Raspberry Pis are also connected to their dedicated AP
with a direct Ethernet cable connection.

\section{Opportunity Evaluation}\label{sec:contention}
The purpose of this section is to quantify what the improvements
are assuming we set an optimal contention window.

We run experiments in our test-bed (see Figure~\ref{testbed}), where we vary the number of APs transmitting data
as well as the contention window used. We also vary the number of APs
we control in the system. The ones we do not control are assumed to follow the
default back-off mechanism (BEB) as well as use the default (EDCA) transmission queue settings.

We measure both the throughput and the latency across all the APs as well as
across APs that we do not control.

We vary the number of APs transmitting between 2 and 8 and the $[cwmin,cwmax]$ between
$[1,1]$ and $[1023,1023]$, and compare this to the default best effort contention window setting of $[15,63]$ as
well as a best effort contention window setting of $[1,1023]$.

The dotted lines in all graphs represent the corresponding value for the
default contention window with exponential backoff (BEB), i.e. for best effort
$[cwmin,cwmax]=[15,63]$ unless otherwise stated. 

\subsection{Throughput Improvement}\label{sec:tp_contention}
\begin{figure}[htbp]
        \centerline{\includegraphics[scale=0.4]{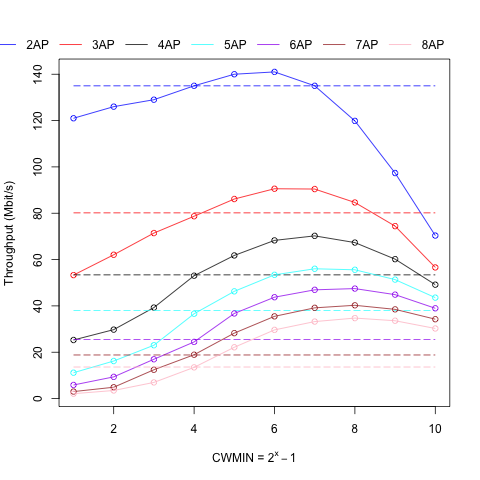}}
        \caption{Median Throughput with different Contention Windows (CW) and transmitting APs.}
\label{contention_tp}
\end{figure}

\textbf{\emph{What is the improvement in throughput when setting an optimal contention window?}}
From Figure~\ref{contention_tp} we see that the median throughput improvement goes up to
155\% with 8 concurrent transmitters when selecting the optimal CW compared to using
the default backoff mechanism.

\subsection{Latency Improvement}
\textbf{\emph{What is the improvement in latency when setting an optimal contention window?}}
We run the same experiments as above but now study the optimal latency in general as well as under
the optimal throughput settings as measured above.

\begin{figure}[htbp]
        \centerline{\includegraphics[scale=0.4]{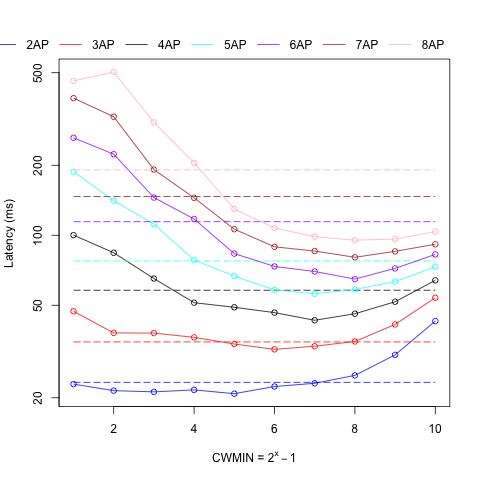}}
        \caption{Median Latency with different Contention Windows (CW) and transmitting APs.}
\label{contention_lt}
\end{figure}

From Figure~\ref{contention_lt} we see that the median latency improvement goes up to
50\% with 8 concurrent transmitters when selecting the optimal CW compared to using
the default backoff mechanism. We also notice that the minimum latency point
tends to coincide with the maximum throughput point (true for 3-8 APs, and close for 2 APs).

\subsection{Partially Managed System}
\textbf{\emph{Can we improve throughput and latency even for streams we do not control?}}
We ran the same experiment as above but we only control 4 of the 8 APs~\footnote{In our setup there is 
a one-to-one mapping between streams and APs, so we could also say we control 4 of 8 streams.}.
The APs we do not control use the default backoff settings and mechanism.

\begin{figure*}
        \centering
        \begin{subfigure}[b]{0.32\textwidth}
                \includegraphics[width=\textwidth]{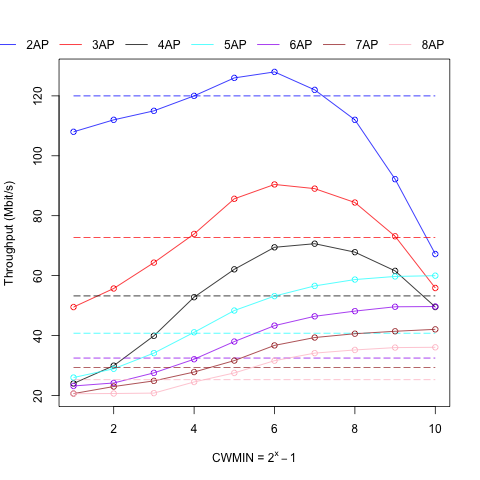}
        \end{subfigure}
        \begin{subfigure}[b]{0.32\textwidth}
                \includegraphics[width=\textwidth]{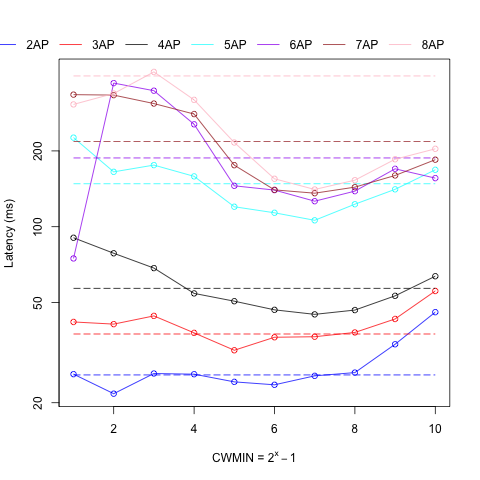}
        \end{subfigure}
        \caption{Overall Throughput and Latency with half of the APs controlled. X axis denotes CWs on APs we control.}
        \label{partialcontrol}
\end{figure*}

Figure~\ref{partialcontrol} shows that
the aggregate throughput and latency still improves when we only control half of the
APs.

\begin{figure*}
        \centering
        \begin{subfigure}[b]{0.32\textwidth}
                \includegraphics[width=\textwidth]{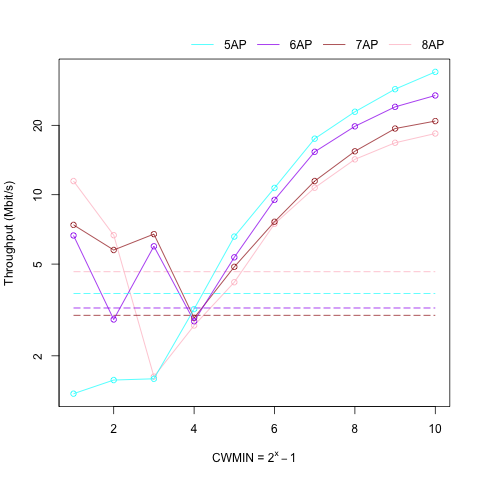}
        \end{subfigure}
        \begin{subfigure}[b]{0.32\textwidth}
                \includegraphics[width=\textwidth]{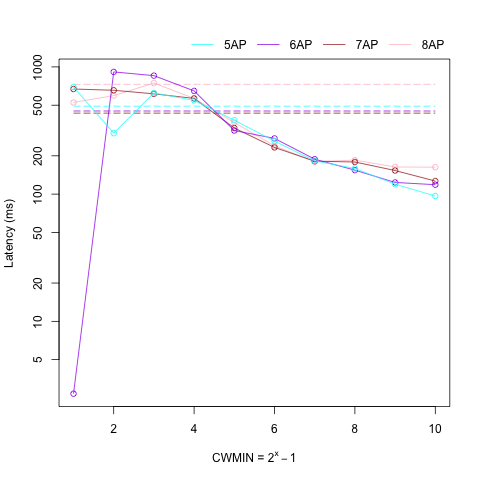}
        \end{subfigure}
	\caption{Throughput and Latency of APs not controlled. X axis denotes CWs on APs we control.}
        \label{partialcontrolnoise}
\end{figure*}

Figure~\ref{partialcontrolnoise} shows that both the throughput and the latency improves
for the APs we do not control. The way to read it is to first 
look at the optimal CW value in Figure~\ref{partialcontrol} for 5, 6, 7 or 8 APs,
and then compare the dotted line to the solid line value for that CW in this figure.

\subsection{Window Range}\label{sec:windowrange}
\textbf{\emph{If we let the contention window be adjusted with the default backoff mechanism (BEB), but
change the minimum and maximum window to be the same as the minimum and maximum windows
used by the dynamic method, does it improve or worsen the throughput and latency?}}~\footnote{note, with
our dynamic method the $CW_{min}$ and $CW_{max}$ are set to the same value, so what we are setting equal 
for BEB and the dynamic method here is the range of all allowable values of CW}

\begin{figure*}
        \centering
        \begin{subfigure}[b]{0.32\textwidth}
                \includegraphics[width=\textwidth]{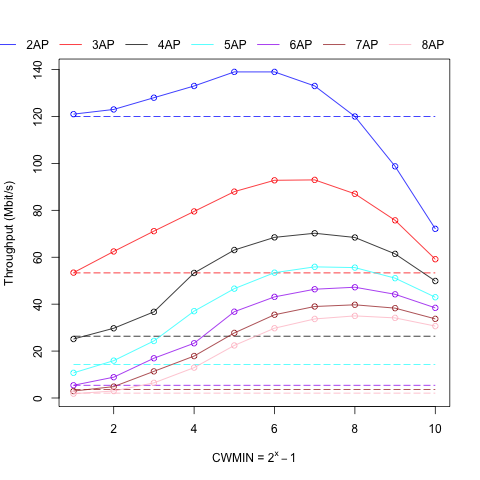}
        \end{subfigure}
        \begin{subfigure}[b]{0.32\textwidth}
                \includegraphics[width=\textwidth]{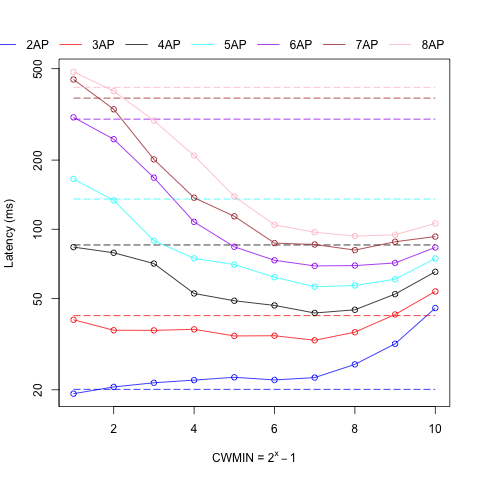}
        \end{subfigure}
        \caption{Throughput and Latency with default cwmin/max [1,1023]}
        \label{windowrange}
\end{figure*}

Note, in the graphs in Figure~\ref{windowrange}, the dotted lines refer to $[cwmin,cwmax]=[1,1023]$. 
If we compare the dotted lines in Figure~\ref{contention_tp} and
Figure~\ref{contention_lt} with the dotted lines in Figure~\ref{windowrange},
we can see that the extended range of the default backoff mechanism has a negative effect
both on the throughput and on the latency. This shows that the default mechanism 
is not good at adjusting to the optimal window. Recall, that the default range for BEB best effort
traffic, which is used here, is $[cwmin,cwmax]=[15,63]$.

\subsection{Collision Reduction with Optimal Contention Window}
\textbf{\emph{What is the reduction in collisions if an optimal CW is used?}}
We collect all the packets sent from 8 concurrent unlimited TCP streams both
in the case of the default back-off mechanism and with the optimal CW for
8 transmitters in a saturated state. We then look at the
802.11 header {\it Frame Control Field} retry bit (12th frame control bit) and count the proportion of packets
that were marked as retries. All packet captures cover 30 seconds (three 10 second bursts).
With an optimal CW the system had $12$\% retries or collisions,
whereas the system with the default mechanism had $61$\%, which is a $78$\% reduction
in collisions (retries). The total packet volume transmitted was $407$ MBytes 
with an optimal CW and $376$ MBytes with default.

\subsection{RTS/CTS Impact}
\textbf{\emph{Does enabling RTS/CTS impact the collision probability?}}
In all the experiments up to this point we have enabled RTS/CTS.
Now, with an optimal CW and RTS/CTS turned off the collision probability measured
as per above was $7$\%, and with the default back-off and no RTS/CTS the
probability of a collision was $33$\%. Hence, in an RTS/CTS scenario
the reduction in collisions was $78$\%. So although there are fewer collisions
overall the proportional reduction is the same. These results are summarized
in Table~\ref{T:collisions}.
The total packet volume transmitted was $406$ MBytes
with an optimal CW and $383$ MBytes with default.
The collision probability drops by half, and the optimal CW also drops slightly, when
RTS/CTS is disabled. More drastically, the throughput (goodput) improvement of setting an
optimal CW goes down to about $13$\% from about $155$\% (see results in Section~\ref{sec:tp_contention} for 8 APs). 
The change is solely due
to the default backoff mechanism improving when RTS/CTS is turned off. The throughput
numbers when using an optimal CW are roughly the same, when RTS/CTS is on and when it is off.
However, RTS/CTS needs to be turned on in dense environments to avoid hidden node
issues, and with sufficiently large number of competing transmitters it will likely
start decaying again even without RTS/CTS.
Note, that this result may seem counter-intuitive as RTS/CTS is often enabled to
decrease collisions. But this experiment shows that there is a point of overload where the RTS/CTS frames
themselves can cause too many collisions.

\begin{table}[htbp]
	\caption{Collision Probability (\%)}
\begin{center}
\begin{tabular}{|c|c|c|}
\hline
 & \textbf{Default} & \textbf{Optimal}\\
\hline
        \textbf{With RTS/CTS} & $61$ & $12$  \\
        \textbf{No RTS/CTS} & $33$ & $7$  \\
\hline
\end{tabular}
\label{T:collisions}
\end{center}
\end{table}

\subsection{Fairness}
\textbf{\emph{Is it fairer to set both cwmin and cwmax to be the same optimal value across all APs than using
an exponential backoff?}}
To test this assertion we run the same benchmark with 8 concurrent transmitters (APs),$n=8$, using unlimited
TCP traffic and measure the Jain Fairness Index~\cite{jain1999}: 
\begin{equation}
	\frac{\left(\sum_{i=1}^n {x_i}\right)^2}{n\sum_{i=i}^n {x_i^2}}
\end{equation}
where $x$ is the throughput and $n$ the number of transmitters.
The value is between $\frac{1}{n}$, minimum fairness, and $1$, maximum fairness.
In a sample of 5 runs, and when setting an optimal cw 
we get a Jain index of $0.982 \pm 0.006$
and for the default backoff mechanism we get $0.980 \pm 0.17$ with a $95$\% confidence band assuming
a normal distribution. The mean value difference is not significant, as both values are very high with 8 streams,
but the variance in values is significantly larger with the default backoff algorithm.
The conclusion is hence that setting the same optimal contention window across all APs is fair, as expected.
We also note that the default $[cwmin,cwmax]$ range for best effort traffic is $[15,63]$ so the unfairness
as a result of doubling the wait time is limited. If we instead set the $[cwmin,cwmax]$ to $[1,1023]$
we get a Jain index of $0.816 \pm 0.059$ which is significantly lower and showcases the danger of
exponential backoff. We also note that Heusse et.al. ran simulations producing similar results in~\cite{heusse2005} (Figure 1).

\section{Model Evaluation}\label{sec:prediction}
We now turn to workload-based experiments where we replay the
data described in Section~\ref{sec:dataset} in our test-bed.
We first analyze the data to determine which type of models
are appropriate, then we discuss our general method in some more detail,
train model meta parameters, and simulate training speed. Finally
we run an online test, benchmarking different models.

\subsection{Model Analysis}
First we investigate the relationship between aggregate load, active
transmitters (APs) and optimal CW. For this purpose we replay 1 hour
(3600 data points) of our 8 workloads concurrently (see Figure~\ref{workloadstreams}). For each data point
replayed we measure the throughput obtained with different
contention windows (1..1023). After the full replay is done we pick the
optimal cw setting given observed throughput from the last period as well
as observed number of active APs. We call this process, {\it offline} or {\it exhaustive} calibration. 

\begin{figure}[htbp]
        \centerline{\includegraphics[scale=0.49]{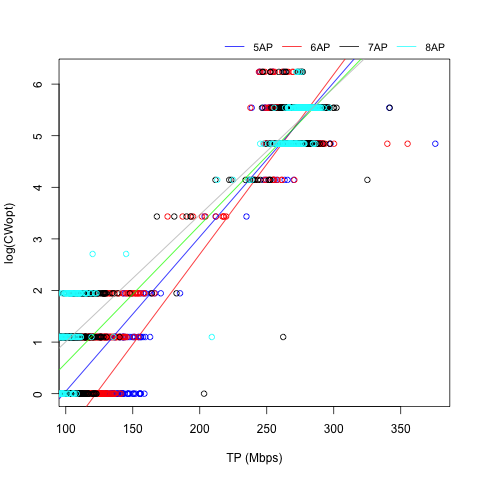}}
        \caption{Offline model fit.}
\label{model}
\end{figure}

We fit Equation~\ref{eq:poly} in Section~\ref{sec:model} to the measurements
and obtain an $R^2$ fit of $0.93$. The fitted model for this hour can be represented as:
\begin{equation}
	log(CW_{opt}) = -3.6 + 0.19 a + 2.7\times10^{-8} tp
\end{equation}
where $t\!p$ is in bits/s. All coefficients are statistically significant.
The regression lines for different
activity levels can be seen in Figure~\ref{model}.
We note that a model without log transform renders an $R^2$ of $0.73$. 

\subsection{Method}
Of course the exhaustive calibration we used above cannot be applied in a real system
where you only get to see each load condition once and only get to pick one contention
window to test. Even if an approximate exhaustive calibration could be performed, we 
want to minimize the training required to fit the parameters and also ensure that the
parameters evolve over time to fit new behavior in the load conditions.
For example a new source of interference introduced could simply be encoded as a new
multiplier in the actives (number of active transmitters) parameter coefficient.

We now describe how to train the models described in Section~\ref{sec:model}, i.e.
how to find and update the coefficients in Equation~\ref{eq:poly} online.
Training of the model proceeds in the following three steps:
\begin{enumerate}
	\item{Create a CW Observation List (CWObs Queue)}
	\item{Update a Table recording optimal CWs (CWMax Table)}
	\item{Fit Regression Model}
\end{enumerate}
The CWObs Queue is comprised of two FIFO sub-queues of observed 4-tuples. One for calibration data and one for predicted values. The 4-tuples include: last observed aggregate throughput ({\it tplast}), last observed active transmitters ({\it actives}), cw
enforced ({\it cwenf}), and throughput obtained with enforced cw ({\it tp}). An example fictive CWObs List can be seen in Table~\ref{T:cwobs_queue}.

\begin{table}[htbp]
	\caption{Example CWObs Queue}
\begin{center}
\begin{tabular}{|c|c|c|c|}
\hline
	\textbf{tplast} & \textbf{actives} &  \textbf{cwenf} & \textbf{tp}\\
\hline
	$123489331$ & $20$ & $15$ & $223489331$  \\
	$223489331$ & $22$ & $31$ & $323489331$  \\
\hline
\end{tabular}
\label{T:cwobs_queue}
\end{center}
\end{table}

The two FIFO sub-queues are needed since calibration involves cycling through all possible
CW values, and if these calibration points disappear the algorithm may be locked into a suboptimal region, if it were solely based on predictions.

The CWMax Table is updated whenever there is new data in the CWObs Queue. The table quantizes {\it actives} into {\it alevel} and {\it tplast} into {\it tlevel} levels and records
the optimal cw {\it cwopt} used for the maximum {\it tp} as {\it mtp} at that level. 
An example fictive CWMax table can be seen in Table~\ref{T:cwmax_table}.

\begin{table}[htbp]
	\caption{Example CWMax Table}
\begin{center}
\begin{tabular}{|c|c|c|c|}
\hline
	\textbf{alevel} & \textbf{tlevel} &  \textbf{mtp} & \textbf{cwopt}\\
\hline
	$1$ & $1$ & $123489331$ & $15$    \\
	$1$ & $2$ & $223489331$ & $31$    \\
\hline
\end{tabular}
\label{T:cwmax_table}
\end{center}
\end{table}

Finally, we fit a ML regression model (see Section~\ref{sec:parameter_estimation}) 
as per Equations~\ref{eq:poly},~\ref{eq:nb_eq}, and~\ref{eq:dnn_eq} with \{{\it alevel,tlevel}\} tuples from the table as features
and {\it cwopt} as targets. Note {\it mtp} is only kept in the table to be able to update {\it cwopt}. 

Now, the model execution (prediction) part of the algorithm can proceed as follows:
\begin{enumerate}
	\item{Map the observed actives and throughput to the corresponding  \{{\it alevel,tlevel}\}-level tuple}
	\item{Use the trained model to predict the {\it cwopt} with this level as input, and set it on all APs}
	\item{Track the cw used and record the throughput obtained in the next time period, and add it to the CWObs Queue}
\end{enumerate}

Note that thanks to the CWobs queue the quantization of levels can be done dynamically based on the current content in the queue, to avoid sparsity
in the CWMax Table. We also need to ensure that the queue has a wide set of {\it cwenf} values, so some initial or 
random exploration at infrequent intervals need to be performed to avoid lock-in into a cw range.

%Conceptually,
%this is similar to $\epsilon$-greedy exploration in reinforcement learning. However, again note that
%we do not have future rewards since we only do single-step predictions, so RL does not apply here.

We also note that the approach assumes that there is a good correlation between the number of active transmitters,
and the aggregate load from one period to the next.  
In other words we assume the process has the  Markov property. 
We have verified (see the ACF charts in Figure~\ref{workloadstreams}) 
that this is true for our traces in periods from 1 to 10s. If this were not true we would not feed the observed
actives and load levels into our models, but the predicted values~\footnote{This is then similar to Q-learning with TD(1)}. We have used some ML models for this task too
but it is overkill with the tested data sets due to the high autocorrelations. For more complex models with hidden
parameters like DNN it could also capture the predictive aspect of these values, but that remains to be verified.

\subsection{Calibration Period Estimation}
To determine how much training data is needed for our 
predictions to reach the optimal performance, and to get
a sense for how frequently the models need to be recalibrated,
we ran training simulations using the data generated by the offline calibration.
Recall, that the offline training data contains a record for each step in the
workload trace for each possible contention window, recording the throughput
obtained.

Simulating a real run thus involves picking a single record corresponding to
a single contention window for each step. We compare
BEB, ABA and MLBA vs an optimal picker that always picks the
contention window with the optimal throughput value.
For MLBA we use Linear Regression (LR), Deep Neural Network (DNN) and 
Naive Bayes (NB) for parameter estimation.

During a training phase we pick random contention windows before we train our model
with the recorded data and start making predictions. Now we vary the steps, recall
each step is a second, used before we start predictions. The predictions are then
used throughout the data set which is one hour, or 3600 data points.

\begin{figure}[htbp]
        \centerline{\includegraphics[scale=0.49]{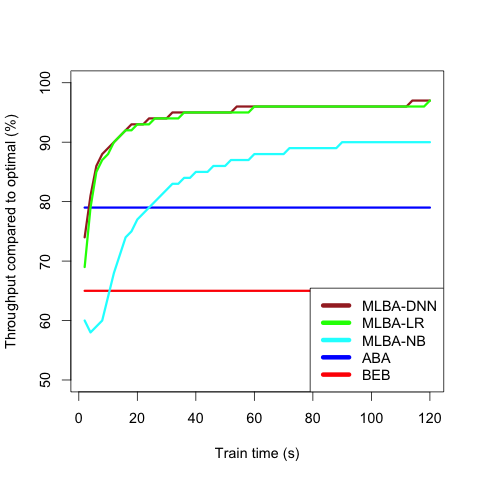}}
        \caption{Prediction performance with different train times.}
\label{train_simulation}
\end{figure}

The throughput obtained compared to the optimal throughput for BEB, ABA and MLBA
are shown in Figure~\ref{train_simulation}.

We note that we reach the performance of ABA with MLB almost instantaneously
with just a few data points in our trained model. The prediction performance
then converges to its best result at about 95\% of optimal after about 35 seconds. 
MLBA-DNN performs marginally better than MLBA-LR and tends to reach plateaus earlier.
MLBA-NB converges very slowly compared to the other MLBA algorithms.
The DNN model is configured with 10 nodes in both hidden layers.

Based on this result we set our calibration period (where we round robin CWs)
to 30 seconds.

\subsection{Training History Estimation}
Now, how much data should we keep in our CWobs and CWmax tables before
we evict old values? And how often should we recalibrate and explore
new CW values that are not predicted to be optimal? To answer those
questions we pick a 15 minute period (900 data points) and run all our training algorithms
with different parameter values for table size and re-calibration probability~\footnote{conceptually,
this is similar to $\epsilon$-greedy exploration in reinforcement learning. However, again note that
we do not have future rewards since we only do single-step predictions, so RL does not apply here.}.
Table~\ref{T:sigsense_summary} shows the results. {\it History} denotes the 
size of the CWobs table and {\it calib.} denotes the probability of
exploring and calibrating new CW values. The table also shows the
median throughput improvement over BEB for the different algorithms.

\begin{table}[htbp]
	\caption{Training History and re-calibration frequency summary}
\begin{center}
\begin{tabular}{|c|c|c|c|c|}
\hline
	\textbf{history} & \textbf{calib. (\%)} & \textbf{LR (\%)} & \textbf{NB (\%)} & \textbf{DNN (\%)} \\
\hline
	$150$ & $1$ & $50.8$ & $54.3$ & $52.9$ \\
	$\mathbf{600}$ & $\mathbf{1}$ & $\mathbf{52.5}$ & $\mathbf{54.1}$ & $\mathbf{55.3}$ \\
	$300$ & $1$ & $54.1$ & $49.5$ & $56.6$ \\
	$300$ & $5$ & $52.7$ & $56.0$ & $51.4$ \\
	$300$ & $10$ & $50.3$ & $43.9$ & $51.1$ \\
\hline
\end{tabular}
\label{T:sigsense_summary}
\end{center}
\end{table}

The configuration with the overall best improvements across all training algorithms was a history of 600 seconds
and re-calibration frequency 1\%~\footnote{i.e., we toss a coin that comes up heads 1\%, 
and if it does, we explore a random CW value instead of the one predicted to be optimal}. 
Hence, we pick that configuration for the subsequent benchmarks. 
We also note that DNN outperformed LR in 4 of 5 tests, and NB in 3 of 5 tests. DNN also achieved the overall
highest improvement, $56.6\%$.

\subsection{Longitudinal Benchmarks}
Finally, using the best performing configuration described above, 
we test our backoff method across five 15 minute-periods succeeding the model building period,
see Figure~\ref{workloadstreams}, and compare the throughput between BEB, ABA, and the MLBA algorithms.
To cover a larger time span, we only test the first 15 minutes of each of the 5 hours in the test set.

\begin{figure}[htbp]
        \centerline{\includegraphics[scale=0.49]{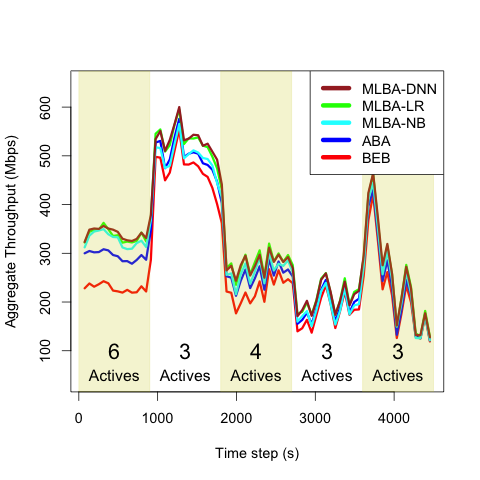}}
	\caption{Prediction test results (1-minute averages) for 5 test periods and with average actives for each period displayed.}
\label{prediction_test}
\end{figure}

The results can be seen in Figure~\ref{prediction_test}.% and are also summarized in Table~\ref{T:prediction_summary}. 
%The online training algorithms described above continuously update their model as the workload trace is replayed.

We group actives into two levels (${0..3}$ and ${4..8}$, $alevel=1$ and $alevel=2$) and use 5 load levels ($tlevel=[0,5)$)
based on measured percentiles to reduce the sparsity of the throughput list data. 
%According to the model building and parameter training
%in the preceeding sections, we start adding training data to the model
%after cycling through 30 data points (round robin over CWs in power of 2 over given range), and start dropping off old training data after 600 points. 
%Note that a point here
%represents a second from the original trace we replay. To avoid being locked into a suboptimal CW range we choose a random CW
%1\% of the time even after the initial training phase. As noted above these calibration values are stored in a different layer of the CWObs queue.
%By keeping the allowed set of CWs small the negative effects of this calibration and exploration can be minimized.
%The DNN model is configured with 10 nodes in both hidden layers.

%\begin{table}[htbp]
%	\caption{Average AP Throughput in Mbps with different models}
%\begin{center}
%\begin{tabular}{|c|c|}
%\hline
% & \textbf{Throughput} \\
% & \textbf{(Mbps)} \\
%\hline
%        BEB & $24.6$  \\
%        ABA & $32.0$ \\
%        MLBA-LR & $36.8$  \\
%        MLBA-DNN & $37.9$ \\
%        MLBA-NB & $37.5$  \\
%\hline
%\end{tabular}
%\label{T:prediction_summary}
%\end{center}
%\end{table}

To test significance of differences between methods we take the empirical percentile of percent difference between the methods, period-by-period.
We then measure the average difference and the percentile where the method tested starts getting a positive difference. For example,
if method A on average across all replayed time steps (seconds) achieves an aggregate throughout across all APs that is 35\% higher than method B
and method A has a higher throughput than B in 70\% of all time steps we say: $Avg(A>B) = 35$ and $SigL(A>B) = 30$. $SigL$ here can be compared to
 a one-tailed statistical significance level, in that a 5\% significance level~\footnote{probability of rejecting null hypothesis $A \leq B$ when true} is equivalent to $SigL(A>B)=5$. Table~\ref{T:sig_summary} summarizes the statistical test results.

\begin{table}[htbp]
	\caption{Average Improvements and Significance Levels. The five time-period results (see Figure~\ref{prediction_test}) are parenthesized.}
\begin{center}
\begin{tabular}{|c|c|c|}
\hline
	\textbf{Hypothesis}& \textbf{Avg (\%)} & \textbf{SigL (\%)} \\
\hline
	ABA > BEB & $13(31,8,14,7,5)$ & $15(\mathbf{5},20,10,20,30)$   \\
	\textbf{LR > BEB} & $\mathbf{25}(49,16,27,17,14)$ & $\mathbf{10}(\mathbf{5},10,10,15,20)$  \\
	\textbf{DNN > BEB} & $\mathbf{26}(53,18,26,16,14)$ & $\mathbf{10}(\mathbf{5},10,10,15,20)$  \\
	NB > BEB & $18(48,10,19,5,8)$ & $20(\mathbf{5},20,15,40,30)$  \\
	LR > ABA & $10(15,8,12,9,9)$ & $15(\mathbf{5},20,15,20,20)$  \\
	DNN > ABA & $11(17,10,11,9,9)$ & $15(\mathbf{5},20,15,20,25)$  \\
	NB > ABA & $4(13,2,5,-2,3)$ & $40(10,50,40,65,45)$  \\
	DNN > LR & $1(3,2,0,-1,0)$ & $50(35,40,55,60,55)$  \\
	DNN > NB & $7(4,9,7,11,6)$ & $25(30,20,25,20,20)$  \\
\hline
\end{tabular}
\label{T:sig_summary}
\end{center}
\end{table}

We note that our MLBA algorithms do better in periods of more heavy load (e.g. period 1 in Table~\ref{T:sig_summary}) compared to both ABA and BEB, 
and those are also the periods where MLBA-DNN tends to do slightly better than MLBA-LR. 
%Next we look at the difference between having one and two levels in our neural network. The one-level hidden network is here referred to as $NN$.
%We find that $Avg(DNN>NN) = 5.0$ and $SigL(DNN>NN) = 20$, i.e. DNN beat NN in 80\% of all time steps replayed.

\section{Implementation}\label{sec:implementation}
We have implemented a patch for hostapd in the latest OpenWrt
release (18.06.1) that exposes the $CW_{min}$ and $CW_{max}$ settings for the transmission
queues. It allows us to control all queues, although our experiments
only change the best effort (BE) queue setting.
The patch uses the hostapd control interface and also adds new hostapd\_cli
commands. 

Depending on the deployment scenario we offer two ways to control the APs.
The LAN controller is appropriate if the controller can be
deployed in the local network, and the Cloud Controller is approprate if
the APs only only had outbound access and should be controlled from
an external network, e.g. a Cloud platform.

\subsection{LAN Controller}
In the LAN controller setup we provide a HTTP REST API exposed from the APs directly. 
Hostapd\_cli is used behind a lighttpd Web server via CGI.
The CGI scripts are then invoked by an HTTP client remotely from the controller.
The roundtrip time to either call our custom CW control APIs or the pre-existing all\_sta
call to collect transmission data is about 20ms, without any optimizations.
To improve the performance further a custom TCP sever embedding the hostapd\_cli protocol
could be used on the AP. The advantage of the current solution is, however, that it
requires minimal modifications on the AP and leverages existing standard applications
already in use in most APs today.

After collecting data with this interface, the controller feeds it into our contention window predictor,
which then uses the load data to estimate active transmitters, and load, and finally predict the optimal CW.
The optimal CW is then set across all APs using the same interface.

Collecting data, and making prediction is all done online and continuously in
windows of 1 to 10s.

The controller and predictor are both written in Python using numpy, sklearn, keras, tensorflow, and scipy, and also comprises a http client that converts the hostapd control
interface output into JSON. The controller also exposes a Flask-based REST interface for monitoring the status of the system and predictions.

\begin{figure}[htbp]
        \centerline{\includegraphics[scale=0.25]{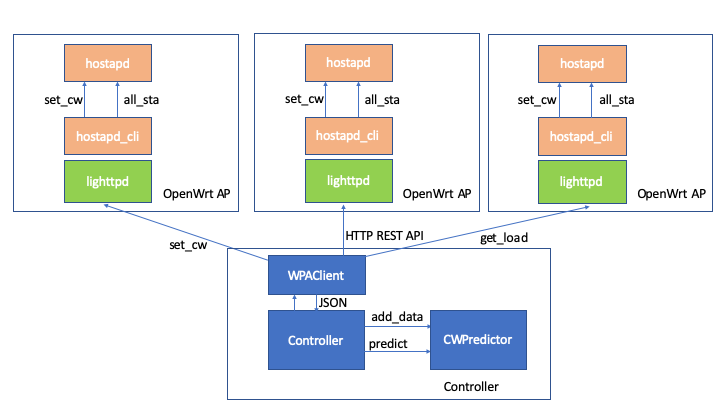}}
        \caption{LAN Controller Contention Window Control Software Architecture.}
\label{lanarch}
\end{figure}

The LAN controller architecture is depicted in Figure~\ref{lanarch}.

All the experiments presented in this paper use the LAN controller API.

\subsection{Cloud Controller}
If the controller that is responsible for changing the contention window and collecting load from the APs is deployed
in a network that does not have inbound access to the network where the APs are deployed, which is typical for Cloud
deployments, we also provide a pubsub API that effectively turns 
pushes into pulls and allows similar control as if the AP API could be accessed
directly. Instead of using the lighttpd server API this model make use of a lua socket client that long-polls a python
Flask server on the remote network. This server in turn has an API that allows direct access to set the contention
window and request load. When load is requested the APs push the current values to a Load cache in the cloud, implemented
using Redis. The controller can then obtain the payload for all APs directly from Redis. Before flushing the load (only
most recent values are kept in memory), the load data is written to a database for off-line analysis. The database
is implemented in InfluxDB. The collection and prediction interval can, like the LAN controller, be set to between 1 to 10s,
note however that this architecture increases the latency both to set the contention window and to collect load, but
what is lost in latency is gained in scalability as all the APs are independently polling the PubSub channel and executing
load and cw change commands. Hence this solution is also recommended for larger deployments.
Finally, we also note that the AP to Controller communication goes across networks and potentially over the Internet, in the
cloud deployment case, so the protocol runs over HTTPS using the OpenWrt luasec package.

\begin{figure}[htbp]
        \centerline{\includegraphics[scale=0.25]{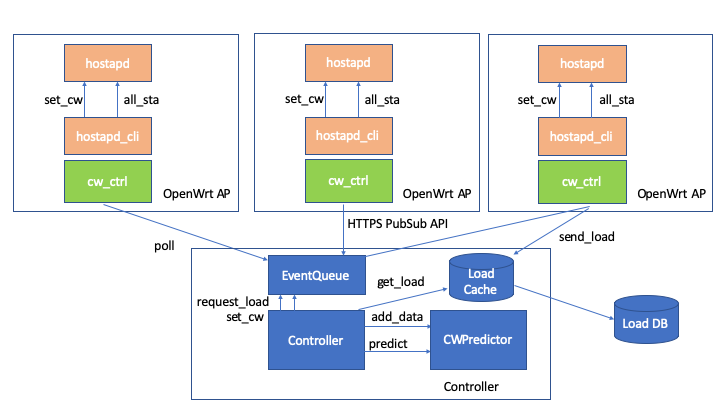}}
        \caption{Cloud Controller Contention Window Control Software Architecture.}
\label{cloudarch}
\end{figure}

The Cloud controller architecture is depicted in Figure~\ref{cloudarch}.

An internal pilot that we currently run with 10 APs is using this architecture.

\section{Related Work}\label{sec:relatedwork}
{\bf Contention window control:} Load and active-transmitter
based adjustment of CW using control theory or theoretically
motivated equations have been studied in~\cite{syed2016,tian2005,xia2006,heusse2005,patras2010,cil2012}.
In addition to applying machine-learning algorithms to capture both active transmitter
and load correlations dynamically, and focussing on AP
control, we also differ from these studies in our focus on test-bed evaluations with off-the-shelf hardware,
and AP control.

{\bf Dense Wi-Fi optimization:} ~\cite{chakraborty2016,murty2008,sheshadri2016,edgewater2018}
propose different load balancing techniques to improve the Wi-Fi QoE dynamically and deal
with interference in dense environments. Our approach differs from this work in that
the control is completely transparent to clients and no new connections have to be re-established
as part of combatting load issues. Wi-Fi 6~\cite{gast2013} offers BSS coloring for spatial reuse,
but only for Wi-Fi 6 clients, and it is based on signal detection thresholds as opposed to
contention window tuning, and is thus orthogonal and complimentary to our approach.  

{\bf Machine-Learning based network management:} In ~\cite{jiang2017,jiang2016,kang2019,xu2018,bega2019,yang2018}
various DNN and Q-Learning approaches are proposed to optimize network configuration. We opted to use
a custom Q-table like structure to train our models and then use an additional step of model fitting, to
allow us to easily test and tune different fitting and online training models separately. Furthermore,
we could not find a good intuitive way to explain why a lower immediate-reward CW could be picked in the current period
just because of a potential future higher reward with another CW, and hence we use single-period predictions. 

{\bf Cognitive Radio control:} A Cognitive Radio Network (CRN)~\cite{haykin2005} shares our goal of improving the efficiency
of a shared spectrum by sensing the environment. The typical goal of a cognitive radio is to determine
whether a primary user (PU) is active on a frequency band, and if not allow secondary users (SU) to use it. Some approaches
take energy sensing data from multiple radios to make a centralized decision~\cite{thilina2013,zhao2006}. Machine learning techniques lend
themselves well to this problem~\cite{clancy2007}, in particular various classification techniques such as 
support vector machines (SVM), k-means clustering, and kernel-based learning have shown good results~\cite{bkassiny2012,ding2013}. Our approach does not require
changing the frequency on any radios, and is thus fully transparent to the receivers, and only require a minimal software
patch on the transmitters.

\section{Conclusion}\label{sec:conclusion}
In this paper we have shown experimentally that there is
a great opportunity to improve upon the 
Wi-Fi EDCA contention window backoff algorithms,
using a predictive model trained with machine learning.

We note that the predictive model is trained for a saturated system.
A non-saturated system does not have the same contention issues, in that
the default BEB algorithm works better under those conditions. 
Hence, a separate learner could be applied to predict whether
the system will be saturated in the next time period, if that is not
the case use the default BEB algorithm.

As a final note, given that we take signals of actual throughput
achieved by the APs we control into account, interfering APs could
to some degree be indirectly accounted for without having to monitor them
directly. The full impact of that is hard to estimate without 
field trials though, which is our next step.

Future work also includes deploying sensors to accommodate more direct load
estimates from exogenous networks, i.e. other Wi-Fi or unlicensed band LTE (e.g. LAA)
traffic not being controlled 
in the prediction of optimal contention window.

\section*{Acknowledgments}
We would like to thank Sayandev Mukherjee for his contributions
to the analytical work providing an intuition behind our model
as outlined in the Appendix.

\bibliographystyle{plain}
\bibliography{related}

\appendix
\section{Appendix}
\subsection{Known results for proportional fair\\ sharing}
Consider $N$ stations, where the transmission probability of station $i$ is denoted $p_i$, $i=1,\dots,N$.  Denote the throughput for station $i$ by $x_i$, $i=1,\dots,N$.  For proportional fairness, we require the optimal transmission probabilities $p_i$ (and thereby the optimal minimum contention windows) to maximize $\sum_{i=1}^N w_i \log x_i$, where $w_i$, $i=1,\dots,N$ are weights for the stations.  In~\cite[Sec.~3.2.1]{siris2006} it is shown that if all stations have the same physical layer transmission rate, then the optimal minimum contention window $\text{CW}_{\min,i}$ of station $i$ to be inversely proportional to the expected throughput $\text{E}[X_i]$ of station $i$:
\begin{equation}
	\frac{1}{\text{CW}_{\min,i}} \propto \text{E}[X_i], \quad i=1,\dots,N.
	\label{eq:PF}
\end{equation}
Further, the expected throughput is given by~\cite[Sec.~3.1]{siris2006}:
\begin{equation}
	\text{E}[X_i] \propto p_i \prod_{\substack{j = 1 \\ j \neq i}}^N (1-p_j), \quad i=1,\dots,N,
\end{equation}
with the transmission probabilities given by
\begin{align}
	p_i &\approx \frac{2}{\text{CW}_{\min,i} + 1} \notag \\
	&\propto \frac{1}{\text{CW}_{\min,i}}, \quad i=1,\dots,N,
	\label{eq:pCW}
\end{align}
and
\begin{equation}
	\text{E}[\text{CW}_i] \approx \text{CW}_{\min, i}, \quad i=1,\dots,N.
\end{equation}

\subsection{Heuristics derived from known results}\label{sec:heuristics}
To derive the heuristic relationship between $\log\,\text{CW}_{\min}$ and $a$, the number of active stations, and $\sum_{j=1}^N x_j$, the throughput, we first assume that $p_i=p$, $i=1,\dots,N$, and $X_1,\dots,X_N$ are independent and identically distributed (i.i.d.). Note that this also implies that  $\text{CW}_{\min,i} = \text{CW}_{\min}$, $i=1,\dots,N$.  Assuming $p$ small and $N$ large, we have
\begin{align}
	\frac{1}{\text{CW}_{\min,i}} &\propto \text{E}[X_i] \notag \\
	 &\propto p_i \prod_{\substack{j = 1 \\ j \neq i}}^N (1-p_j) \notag \\
	 &\approx p \exp[-(N-1)p] \notag \\
	 &\approx p \exp(-Np) \notag \\
	 &\approx \frac{2}{\text{E}[\text{CW}_{\min,i}] + 1} \exp(-Np) \notag \\
	 &\propto \exp(-Np).
	 \label{eq:heuristic1}
\end{align}

\subsubsection{Heuristic relationship between $\log\,\text{CW}_{\min}$ and $a$}
Note that $Np$ is the mean number of active stations, i.e., it is $\text{E}[A]$, where $A$ is the random variable representing the number of active stations.  If we focus on a sample size of one (in the transmission interval under consideration), we substitute the observed quantity $a$ for the mean $\text{E}[A] = Np$ in~\eqref{eq:heuristic1}, thereby yielding
\begin{equation}
	\frac{1}{\text{CW}_{\min,i}} \propto \exp(-a)  \Leftrightarrow \log\,\text{CW}_{\min} \propto a.
	\label{eq:heuristic_a}
\end{equation}

\subsubsection{Heuristic relationship between $\log\,\text{CW}_{\min}$ and $\sum_{j=1}^N x_j$}
From the inverse relationship between $p$ and $\text{CW}_{\min}$ given by~\eqref{eq:pCW}, and the inverse relationship between $\text{CW}_{\min}$ and $\text{E}[X_i]$ given by~\eqref{eq:PF}, we can rewrite $Np$ as follows:
\begin{align}
	Np &\propto \frac{N}{\text{CW}_{\min}} \notag \\
	&\propto N\,\text{E}[X_i] \notag \\
	&= \text{E}\left[\sum_{j=1}^N X_j\right],
\end{align}
which when combined with~\eqref{eq:heuristic1} yields
\begin{equation}
	\frac{1}{\text{CW}_{\min,i}} \propto \exp\left\{-\text{E}\left[\sum_{j=1}^N X_j\right]\right\}.
	\label{eq:heuristic2}
\end{equation}
Again, focusing on a sample size of one (in the transmission interval under consideration), we substitute the observed quantity $\text{tp} \equiv \sum_{j=1}^N x_j$ for the mean $\text{E}[\sum_{j=1}^N X_j]$ in~\eqref{eq:heuristic2}, thereby yielding
\begin{equation}
	\frac{1}{\text{CW}_{\min,i}} \propto \exp\left(-\sum_{j=1}^N x_j\right) \Leftrightarrow \log\,\text{CW}_{\min} \propto \text{tp}.
	\label{eq:heuristic_tp}
\end{equation}

%%%%%%%%%%%%%%%%%%%%%%%%%%%%%%%%%%%%%%%%%%%%%%%%%%%%%%%%%%%%%%%%%%%%%%%%%%%%%%%%
\end{document}